\documentclass[12pt,nofootinbib]{revtex4-2}

\usepackage[utf8]{inputenc}
\usepackage[english]{babel}
\usepackage{mathtools}
\usepackage{amsmath}
\usepackage{amsfonts}
\usepackage{amssymb}
\usepackage{amsthm}
\usepackage{suffix}
\usepackage{physics} 

\usepackage{hyperref} 
\usepackage{subcaption} 
\usepackage{comment} 
\usepackage{float} 
\theoremstyle{plain}
\newtheorem{thm}{Theorem}[section]
\newtheorem{prop}[thm]{Proposition}

\theoremstyle{definition}

\newcommand*{\myproofname}{Proof} 
\newenvironment{mainproof}[1][\myproofname]{\begin{proof}[#1]
}{\end{proof}}

\usepackage{xcolor} 

\usepackage{bm} 
\renewcommand{\it}[1]{\textit{#1}}
\renewcommand{\bf}[1]{\textbf{#1}}
\newcommand{\up}[1]{\mathrm{#1}}

\renewcommand{\a}{\alpha}
\renewcommand{\b}{\beta}
\newcommand{\g}{\gamma}
\renewcommand{\d}{\delta}
\renewcommand{\th}{\theta}

\renewcommand{\k}{\kappa}
\renewcommand{\r}{\rho}

\renewcommand{\t}{\tau}
\newcommand{\w}{\omega} 

\newcommand{\G}{\Gamma}

\newcommand{\W}{\Omega} 

\usepackage{bbm} 
\newcommand{\bb}[1]{\mathbb{#1}} 

\newcommand{\Z}{\bb{Z}} 
\newcommand{\R}{\bb{R}} 

\newcommand{\DeclareAutoPairedDelimiter}[3]{%
  \expandafter\DeclarePairedDelimiter\csname Auto\string#1\endcsname{#2}{#3}%
  \begingroup\edef\x{\endgroup
    \noexpand\DeclareRobustCommand{\noexpand#1}{%
      \expandafter\noexpand\csname Auto\string#1\endcsname*}}%
  \x}
\DeclareAutoPairedDelimiter{\p}{(}{)} 

\newcommand{\f}[2]{\frac{#1}{#2}} 
\newcommand{\too}{\longrightarrow} 
\let\baraccent=\= 
\renewcommand{\=}[1]{\stackrel{#1}{=}} 

\renewcommand{\bar}[1]{\mkern1mu\overline{\mkern-1mu#1\mkern-1mu}\mkern1mu}
\renewcommand{\tilde}[1]{\widetilde{#1}}
\let\cross\relax 
\newcommand{\cross}{\times}

\begin{document}
\title{The Tortoise and the Hare: \\ A Causality Puzzle in AdS/CFT}
\author{David Berenstein and David Grabovsky}
\address{Department of Physics, University of California, Santa Barbara, CA 93106}
\date{\today}

\begin{abstract}
We pose and resolve a holographic puzzle regarding an apparent violation of causality in AdS/CFT. If a point in the bulk of $\mathrm{AdS}$ moves at the speed of light, the boundary subregion that encodes it may need to move superluminally to keep up. With $\mathrm{AdS}_3$ as our main example, we prove that the finite extent of the encoding regions prevents a paradox. We show that the length of the minimal-size encoding interval gives rise to a tortoise coordinate on $\mathrm{AdS}$ that measures the nonlocality of the encoding. We use this coordinate to explore circular and radial motion in the bulk before passing to the analysis of bulk null geodesics. For these null geodesics, there is always a critical encoding where the possible violation of causality is barely avoided. We show that in any other encoding, the possible violation is subcritical. 
\end{abstract}

\maketitle
\newpage

\section{Introduction}
\label{sec:intro}

\paragraph*{Invitation.}
Imagine that you are wandering a circular room illuminated only by a lamp in its center. Your shadow dances on the walls as you move, by turns larger as you near the center and smaller as you near the edge. If you move at the speed of light, your shadow will have to move even faster to keep up with you.\footnote{Apparent faster-than-light travel is common in the observation of quasar jets in astrophysics \cite{Rees,zensus_pearson_1988}. A detailed understanding of the kinematics of light emission by matter---when, from where, and in which direction they are emitted and observed---resolves the apparent paradox here.} Ordinarily this is not a problem, since shadows carry no physical information. But this is a metaphor for holographic duality, where both you and your shadows are real. The room is anti-de Sitter (AdS) space, the walls are home to a conformal field theory (CFT) on its boundary, and you are a local bulk field configuration---a particle---encoded nonlocally on the walls. An apparent paradox emerges: lightspeed motion in the bulk seems to violate causality on the boundary!

\paragraph*{AdS/CFT and entanglement wedges.} The AdS/CFT correspondence states that a semiclassical theory of quantum gravity in an asymptotically AdS spacetime is equivalent to a conformal field theory on the conformal boundary of that spacetime \cite{Maldacena:1997re,Witten:1998qj,Gubser:1998bc}. Even if one only has access to the quantum information contained in a subregion of the boundary, one can completely reconstruct a corresponding part of the bulk theory. This is usually exemplified by the HKLL construction \cite{Hamilton:2006az} (see also \cite{Banks:1998dd} for earlier work on this). More precisely, suppose that all of the local and nonlocal CFT operators supported on some subregion of the boundary are given. Then there is a region of AdS, called the entanglement wedge of the boundary subregion, within which one can realize any bulk field as a nonlocal CFT operator smeared over the boundary subregion. 

For static bulk backgrounds, there are canonical constant-time slices orthogonal to a timelike Killing field. In this case, the entanglement wedge is found using the Ryu-Takayanagi prescription \cite{Ryu:2006bv,Ryu:2006ef}: one studies the minimal-area surface in the bulk, anchored on the boundary of the region, which is homologous to it and lives in the given constant-time slice. The entanglement wedge is then the region bounded by this surface and the boundary.\footnote{A more general prescription is available for dynamical geometries \cite{Hubeny:2007xt}, but will not be necessary here.} For pure states, the boundary subregion complementary to a given one ends up encoding the corresponding complementary region of the bulk \cite{Dong:2016eik}. There are some additional restrictions, in that the operators supported on the boundary region cannot cause a large backreaction in the bulk: this is part of the idea that the reconstruction occurs only in a code subspace \cite{Almheiri:2014lwa}.

\begin{figure}[ht]
    \centering
    \begin{minipage}{.49\textwidth}
        \centering
        \includegraphics[width=.8\linewidth]{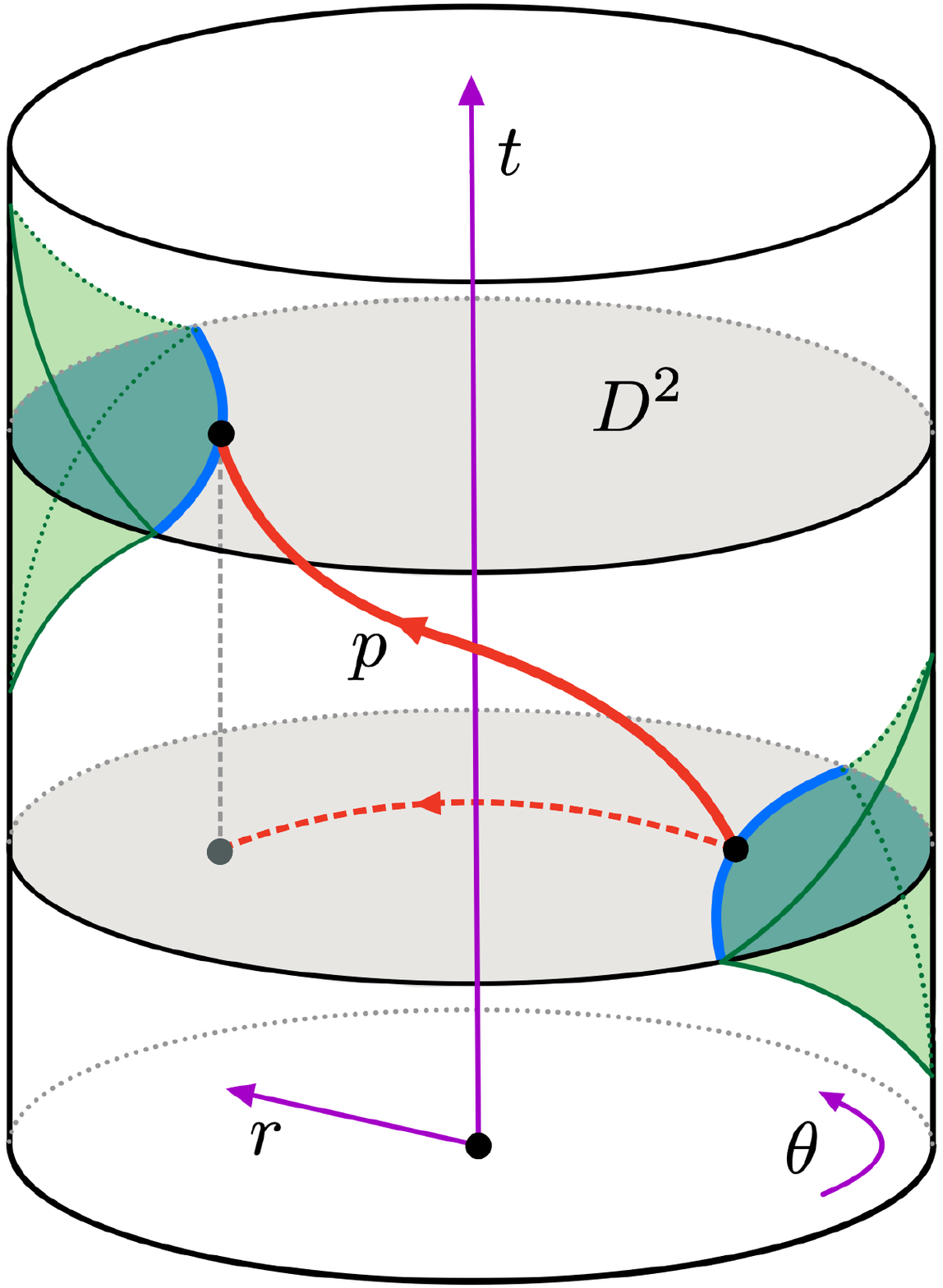}
    \end{minipage}%
    \begin{minipage}{.51\textwidth}
        \centering
        \includegraphics[trim = 140 230 120 215, clip,
        width=\linewidth]{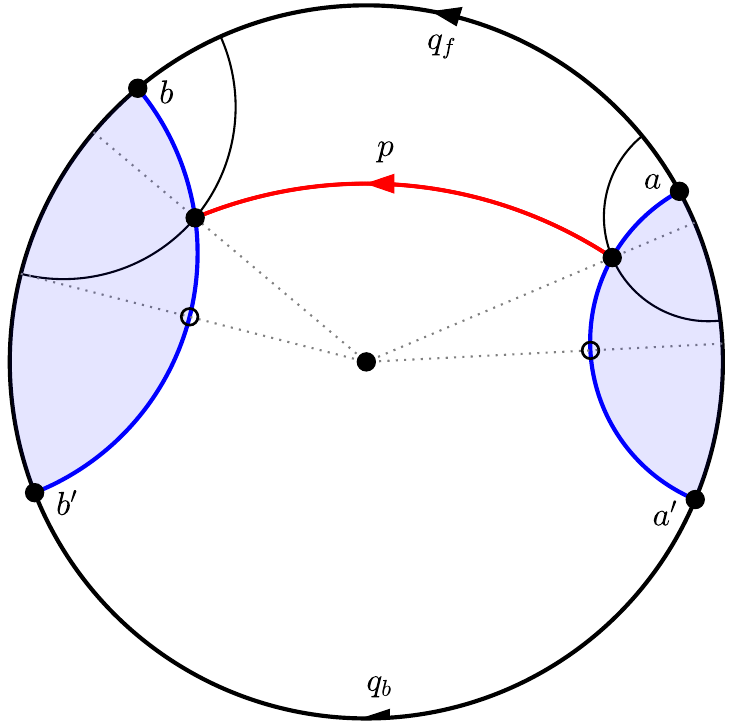}
    \end{minipage}
    \caption{Left: a bulk point $p$ moves in $\up{AdS}_3$. Its encoding on the boundary and the corresponding entanglement wedges are shown on two different constant-time slices. ${}$ Right: the race between $p$ and $q$ for generic paths and non-minimal encoding intervals.}
    \label{fig:introduction}
\end{figure}

This has a particularly simple realization in pure $\up{AdS}_3$, which can be visualized as a solid cylinder whose constant-time slices are Poincar\'e disks $D^2$. The geodesics of $D^2$, shown in Fig. \ref{fig:introduction}, are circles orthogonal to its boundary.\footnote{This statement holds in a coordinate system where the metric on a given constant-time slice is conformally equivalent to the flat metric in a flat coordinate system.} Each geodesic subtends an interval of the boundary circle, which we call the \it{encoding interval}, and is uniquely labeled by its ``deepest'' bulk point, which we call its \it{gem}. As we enlarge the encoding interval, its gem moves further towards the center of the disk. The constant-time slice of the entanglement wedge of a given encoding interval, shaded in Fig. \ref{fig:introduction}, is the region of $D^2$ enclosed by the boundary of $D^2$ and the subtending geodesic. We regard the geodesic itself as part of the entanglement wedge: if we enlarge the encoding interval infinitesimally, the points on the geodesic will lie in the entanglement wedges of both the encoding interval and its complement (this follows the ``all-or-nothing'' point of view of \cite{Engelhardt:2017iwl}). Each boundary interval encodes all bulk points $p \in D^2$ caught in its entanglement wedge, and conversely, $p$ is encoded on any boundary interval that has $p$ in its entanglement wedge. Among these intervals is a unique one of minimal length. Its gem is $p$ itself, and it provides the ``most efficient,'' or \it{minimal}, encoding of $p$.

\paragraph*{A causal puzzle.}
Our goal in this paper is to better understand causality from the perspective of encodings. This turns out to be very nontrivial, as we explain presently. The null motion of a local bulk field in AdS may drag (the midpoint of) its encoding interval around the boundary faster than light. This is easiest to visualize in another geometry: AdS black holes. There one can show that circular timelike geodesics in the bulk have angular velocities greater than one \cite{Berenstein:2020vlp}, the speed of light on the boundary. Finding paradoxical configurations takes slightly more effort in global AdS. Na\"ively, null motion in AdS could cause the initial and final encoding regions to become spacelike-separated, causing the operators supported on each boundary region to mutually commute. The bulk point would be unable to carry information from one interval to the other. To carry information, an action must have a nontrivial response. In linear response theory, one needs a non-vanishing retarded Green's function, computed by a commutator between operators in the initial and final boundary regions. If all of the commutators vanish by boundary causality, but the motion of the bulk point is our means of conveying a message, then we have a contradiction. We can make matters worse by using an encoding where the bulk point starts near the edge of its encoding interval in its direction of motion, but ends up near the opposite edge when it arrives. Making the intervals asymmetric (i.e. non-minimal) and having them ``swing around'' can, in principle, exacerbate the problem.

We can only resolve this puzzle if the encoding intervals are ``big enough.'' As shown in Fig. \ref{fig:introduction}, the instant that $p$ departs from its starting point in the bulk, we send a light signal $q$ along the boundary, starting in the initial encoding region and moving at speed $c=1$ (by convention). Like a photophobic ant hiding in the shadows on the wall, the signal must always stay inside the (moving) encoding interval if causality is to be preserved. The signal must travel from the ``leading'' edge $a$ of the initial interval to the ``trailing'' edge $b$ of the final interval before the bulk point arrives; the tightest bounds are obtained if $p$ lies on the geodesic defining its encoding interval instead of strictly inside. Due to the possible asymmetry of the encoding interval about the bulk point, the ``forward'' signal ($a \too b$) might fail to preserve causality. Instead, the encoding intervals must bow out far enough to let a ``backward'' signal ($a' \too b'$) preserve causality whenever the forward signal does not. 

It is worth noting that our investigation bears some similarity to the problem of gravitational time delay studied by Gao and Wald \cite{Gao:2000ga, Witten:2019qhl}. A special case of the Gao-Wald theorem guarantees that for any causal bulk path that starts and ends on the boundary of empty AdS, a light signal traveling entirely on the boundary will always beat the bulk point. Our purpose here is to slightly extend this instance of the Gao-Wald result to motion deep in the bulk. The discussion above provides a prescription for which boundary paths to compare to the bulk path, and this paper seeks to prove that such a statement holds in empty $\up{AdS}$. More broadly, we expect that this generalized  causality will be preserved if the null energy condition is satisfied in the bulk, even for dynamical gravity \cite{Headrick:2014cta}. In this paper we elaborate on the example illustrated by Figure 1 of \cite{Headrick:2014cta}, which shows that the RT formula marginally satisfies causality on a plane generated by null geodesics. Here we choose different entanglement wedge encodings for the bulk point, and then check how close we come to a critical situation where the bulk and the boundary travel times coincide. The main point is that we can be very explicit: the computations are done directly using elementary geometry. This should, in principle, also inform us better as to how the different code subspaces of the bulk theory talk to each other.

\paragraph*{Outline.} In this paper, we resolve the paradox in several special cases before giving a general proof. In section \ref{sec:illus}, we bring the paradox into stark relief by studying circular and radial bulk motion; along the way, we describe the geometry of encoding intervals and prove that their length acts as an infalling tortoise coordinate on $\up{AdS}_3$. In section \ref{sec:coordinate}, we prepare the ground for the general case: we first parametrize null geodesics, the fastest possible bulk paths, using the tortoise coordinate; then, we discuss the minimality of encoding intervals in more detail. In section \ref{sec:caus} we prove, in several settings of increasing generality, that the bulk point's travel time always exceeds the shortest of the forward and backward signals' travel times. We also explain how our main result and proof is valid in $\up{AdS}_d$ with small modifications, as long as we use spherical regions in the boundary as a substitute for encoding intervals. 
We conclude in section \ref{sec:disc} with a discussion of how tight the bounds on causality are, and discuss generalizations to higher dimensions and more general spacetimes.

\section{An Illustration of Causality}
\label{sec:illus}

We begin by putting a particle (what we call a bulk field) in a rocket ship and sending it on circular and radial trips through $\up{AdS}_3$. In both cases we force the rocket to move along a null path, and we encode the bulk point minimally on the boundary. Even in these simplest examples, the encoding region moves superluminally, making the paradox apparent.

\subsection{Circular Motion}

\begin{figure}[H]
    \centering
    \includegraphics[trim = 140 230 120 215, clip, 
    width=0.57\linewidth]{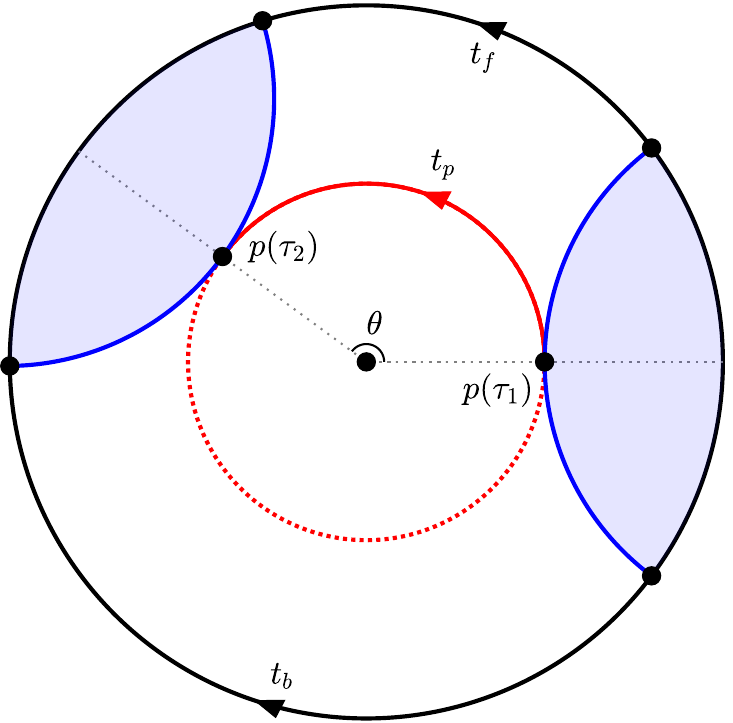}
    \caption{Circular motion in the bulk.}
    \label{fig:toy_model_circ}
\end{figure}

$\up{AdS}_3$ has topology $\R \times D^2$, where $\R$ is the time axis with coordinate $t$ and $D^2$ is the Poincar\'e disk. Putting (Euclidean) polar coordinates $(r,\th)$ on $D^2$ gives $\up{AdS}_3$ the metric
\begin{align}
\label{eqn:AdS_metric0}
\dd s^2_{\up{AdS}_3} = -\p{\f{1+r^2}{1-r^2}}^2 \dd t^2 + 
\f{4\p{\dd r^2 + r^2\, \dd\th^2}}{\p{1-r^2}^2} = 
\dd s^2_{\R} + \dd s^2_{D^2}.
\end{align}
These are sometimes called ``sausage coordinates'' on AdS \cite{Aminneborg:1997pz, Bengtsson:2014fha}, and in these coordinates $D^2$ is conformal to a flat disk in polar coordinates. In sausage coordinates, the geodesics of $D^2$ are arcs of circles that meet the boundary at right angles. The coordinate $r$ should not be confused with with the standard global AdS coordinate $\r$, which measures the distance to the origin in $D^2$. Fixing the $r$-coordinate of the bulk point $p$, let $\g(p) \in [0,\f{\pi}{2}]$ denote half of the (Euclidean) arclength of its minimal encoding interval, as shown in Fig. \ref{fig:toy_model_circ}. 

Now we send $p$ along a null, circular worldline, so that $\dd r = 0$; hence $\g(p)$ remains constant throughout the trip. We show in section \ref{sec:tortoise} that $\w \equiv \abs{\dv{\th}{t}} = \f{1}{\cos\g} > 1.$ Unless $p$ moves along the boundary ($r=1$) or sits motionless at the origin ($r=0$), the center point of its encoding interval always seems to move faster than light. 

To resolve the paradox, we compare the bulk travel time $t_p$ to that of the signals described in section \ref{sec:intro}. If $p$ moves through angle $\th$, its transit takes time $t_p = \f{\th}{\w} = \th \cos\g$. Meanwhile, the signals' travel times are exactly equal to the angles they traverse. Referring to Fig. \ref{fig:toy_model_circ}, the forward signal takes time $t_f = \th - 2\g$, while the backward signal takes time $t_b = 2\pi - \th - 2\g$. (Note that it is no loss of generality to treat the motion in a single constant-time slice, since both the bulk point and the boundary signal move uniformly in $t$.) We now prove that at least one of the signals always beats $p$ to its destination.

\begin{prop}
For all $\th \in [0,2\pi)$ and $\g \in [0,\f{\pi}{2}]$, $\Delta t \equiv \max\{\Delta t_f, \Delta t_b\} \geq 0$, where
\begin{align}
\begin{aligned}
\Delta t_f &\equiv t_p - t_f = 2\g - \th(1 - \cos\g), \\ 
\Delta t_b &\equiv t_p - t_b = 2\g + \th(1 + \cos\g) - 2\pi.
\end{aligned}
\end{align}
\end{prop}

\begin{mainproof}
We begin by noting that both $-\Delta t_f$ and $-\Delta t_b$ are convex functions of $\g$. Indeed, consider their second derivatives: $-\Delta t_f''(\g) = -\Delta t_b''(\g) = \th \cos\g \geq 0$. It remains to ``anchor down'' $\Delta t_f$ and $\Delta t_b$. When $\th \in [0,\pi]$, $\Delta t_f(0) = 0$ and $\Delta t_f(\f{\pi}{2}) = \pi - \th \geq 0$. Thus $\Delta t_f \geq 0$ for $\th \in [0,\pi]$ by convexity and the intermediate value theorem. A similar argument applies to $\Delta t_b$: when $\th \in [\pi,2\pi)$, $\Delta t_b(0) = 2\th - 2\pi \geq 0$ and $\Delta t_b(\f{\pi}{2}) = \th - \pi \geq 0$; hence $\Delta t_b \geq 0$ for $\th \in [\pi,2\pi]$. (For $\th = \pi$, we have $t_f = \pi - 2\g = t_b \leq t_p = \pi \cos\g$.)
\end{mainproof}

We have proven that the forward signal always beats the bulk point on short trips, that the backward signal always wins on long trips, and moreover that they trade off at $\th = \pi$, where $\Delta t = t_f-t_b = 0$.  Notice also that $\w$ becomes arbitrarily large if the bulk point rotates at the center of $D^2$. Here $p$ is stationary, but its (minimal) encoding interval covers half of the boundary circle and rotates with formally infinite angular velocity, its size preventing a violation of causality. In this case the travel time on the boundary and in the bulk are identical (they both vanish), so there is no room for error. We call such situations {\em critical}.

Apparent motion faster than light also occurs when one studies the phase velocity of a wave. Stretching the analogy, we might imagine the midpoint of the entanglement wedge as moving with the phase velocity, while the time of flight between the intervals' endpoints measures the group velocity, in this case the speed of light. This analogy is somewhat strained: as we discuss in detail in the rest of the paper, we have a good deal of freedom in choosing entanglement wedges. There is no canonical notion of phase velocity independent of the choices we make, owing essentially to the nonlocality of the bulk encoding.

\subsection{The Tortoise Coordinate}
\label{sec:tortoise}

The calculation above demonstrates that the resolution of the paradox relies crucially on the finite extent of $\g(p)$. Evidently $\g$ is simultaneously a measure of (1) the size of the encoding interval of $p$, (2) the nonlocality of its encoding, (3) the ``danger'' $p$ poses to causality, and (4) the depth of $p$ in the bulk. To explore this last observation, it will be convenient to study $D^2 \subset \R^2$ in terms of Euclidean plane geometry.

\begin{figure}[H]
\centering
\includegraphics[trim = 120 230 95 215, clip,
width=0.57\textwidth]{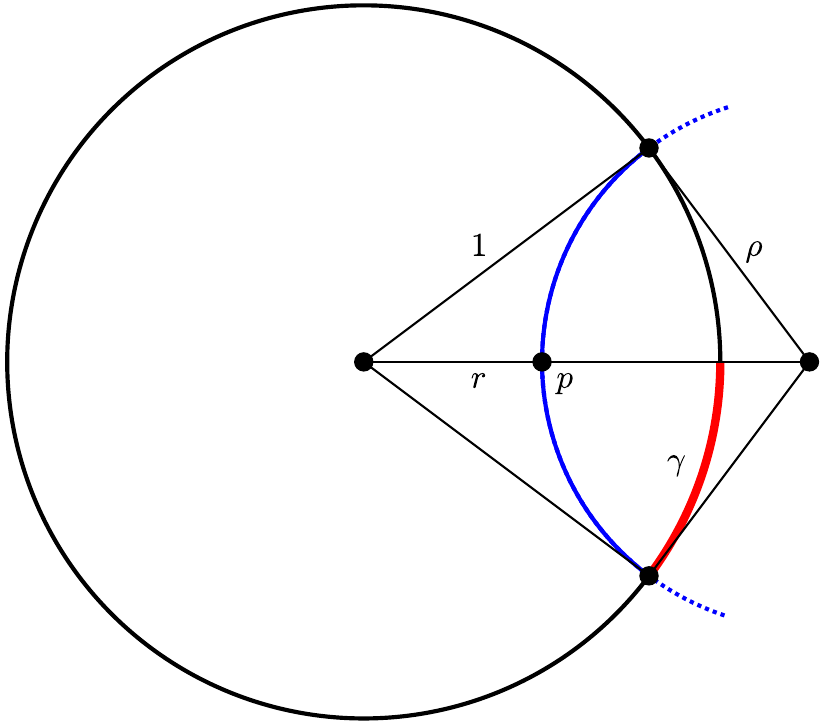}
\caption{The construction of encoding intervals, and the relation between $r$ and $\g$.}
\label{fig:geometry_1}
\end{figure}

Notice that $\g(p)$ depends only on the radial distance $r(p)$ of $p$ from the origin. As $r \too 0$, we have $\g \too \f{\pi}{2}$; as $r \too 1$, we have $\g \too 0$. Thus inside $D^2$, $\g$ parametrizes the radial position of $p$. To relate $\g$ to $r$, we let $\r$ be the radius of the orthogonal circle defining the the minimal geodesic through $p$, as shown in Fig. \ref{fig:geometry_1}. The Pythagorean theorem gives
\begin{align}
\label{eq:new_param}
\r^2 + 1 = (\r + r)^2 \implies \r = \f{1-r^2}{2r},
\end{align}
Since $\g$ is half of the angle subtended by this arc, we find
\begin{align}
\label{eqn:trig}
\sin\g = \f{\rho}{\rho + r} = 
\f{1-r^2}{1+r^2}, \qquad
\cos\g = \f{1}{\rho + r} = \f{2r}{1+r^2}, \qquad
\tan\g = \rho = \f{1-r^2}{2r}.
\end{align}
All three of these relations will prove useful. The last one yields a quadratic equation relating $r$ and $\g$, of which we take the positive root to ensure that $r > 0$:
\begin{align}
\label{eqn:r_to_gamma}
1 - r^2 = 2r \tan\g \implies r = \sec\g - \tan\g.
\end{align}
Substituting the change of coordinates (\ref{eqn:r_to_gamma}) in the metric (\ref{eqn:AdS_metric0}), we rewrite the $\up{AdS}_3$ metric:
\begin{align}
\label{eqn:AdS_metric}
\dd s^2_{\up{AdS}_3} = -\p{\f{1+r^2}{1-r^2}}^2 \dd t^2 + 
\f{4\p{\dd r^2 + r^2\, \dd\th^2}}{\p{1-r^2}^2} =
\f{1}{\sin^2\g} \Big(\!-\dd t^2 + \dd\g^2 + \cos^2\g\, \dd\th^2\Big).
\end{align}
These new coordinates explain the angular velocity quoted above: if $p$ stays at fixed radius ($\dd\g = 0$) and has null trajectory ($\dd s^2 = 0$), then we find $\dd t = \pm \cos\g\, \dd\th$, so $\w = \f{1}{\cos\g}$.

\subsection{Radial Infall}
\label{subsec:radial}

The physical meaning of $\g$ is revealed when we consider the metric induced on a radial ($\dd\th = 0$) null worldline. Since $\g$ and $t$ enter into $\dd s^2$ with no relative factors,
\begin{align}
\dd s^2 = 0 \implies \dd t = \pm \dd \g \implies \g - \g^* = \pm t.
\end{align}
We interpret $\g$ as an infalling Eddington-Finkelstein, or tortoise, coordinate \cite{Wald:GR}. This analysis extends to AdS spacetimes of any dimension, where one needs a higher-dimensional analog of the orthogonal circles in $D^2$. If we use spherical regions of the boundary as substitutes for encoding intervals, then the minimal surfaces anchored on the boundary will be sphere caps orthogonal to the boundary: see for example \cite{Berenstein:1998ij}.

\begin{figure}[H]
    \centering
    \includegraphics[trim = 140 230 120 215, clip,
    width=0.57\linewidth]{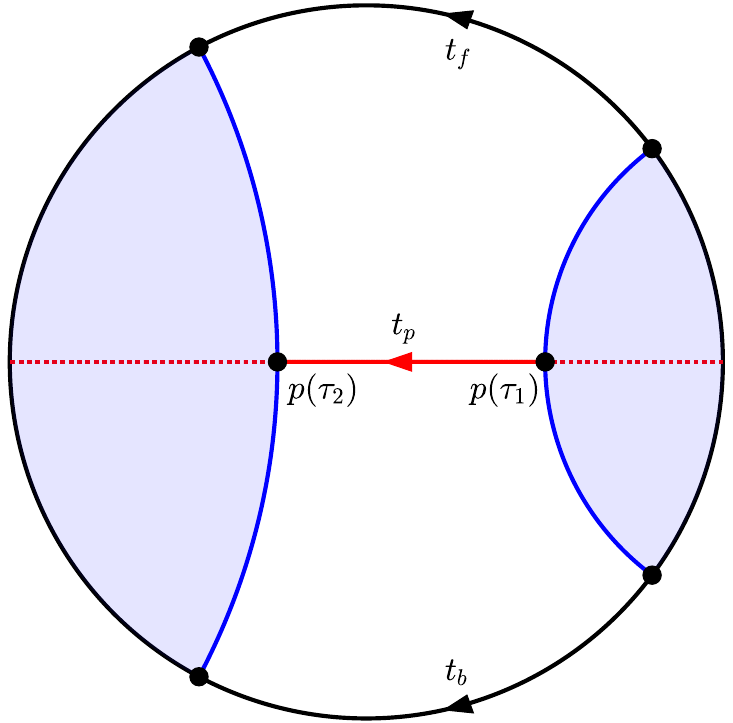}
    \caption{Radial motion in the bulk.}
    \label{fig:toy_model_rad}
\end{figure}

Let us examine radial null paths, shown in Fig. \ref{fig:toy_model_rad}. The null-worldline condition gives
\begin{align}
\dd s_p^2 = \f{1}{\sin^2\g} \p{-\dd t^2 + \dd\g^2} = 0 \implies \dd t = \pm \dd\g \implies v \equiv \abs{\dv{\g}{t}} = 1.
\end{align}
The unit speed of light is expected: the purpose of $\g$ is to parametrize radial lightrays as if they were signals. As $p$ travels into the bulk, $\g$ expands until $p$ reaches the origin. As it crosses the origin, the encoding interval---which occupies half of the boundary---discontinuously jumps to occupy the opposite half and begins to shrink. Even though the encoding interval moves with infinite speed when $\g(p) = \f{\pi}{2}$, its size again prevents a paradox. Regardless of whether $p$ crosses the origin, its travel time matches that of both signals:
\begin{align}
\begin{aligned}
t_p(\g_1 \too \g_2) &= \g_2 - \g_1 \!\qquad = t_f = t_b,
\qquad (p\; \text{does not cross the origin}), \\
t_p(\g_1 \to O \to \g_2) & = \pi - \g_2 - \g_1 = t_f = t_b, 
\qquad (p\; \text{does cross the origin}).
\end{aligned}
\end{align}

In both cases, the time delay vanishes identically. Both the forward and backward signals just barely preserve the causality, independent of the initial and final positions $\g_1, \g_2$. Thus the encoding of $p$ is critical throughout its trip through the bulk.

The examples above illustrate that roughly speaking, $\g$ measures causality in the bulk, while $\th$ measures causality on the boundary. The example of circular motion introduces trade-offs between the forward and backward signals, while the example of radial infall forces us to extend $\g$ beyond the origin, and to confront the discontinuity of encoding there.

\section{Coordinate Gymnastics}
\label{sec:coordinate}

Let us strategize. The examples above fail to be general on two counts: (1) circular and radial paths are not generic, and the circular paths are ``too slow'' in the bulk; and (2) the bulk encoding need not be minimal. To remedy (1), we should study bulk motion along geodesics; in particular, we will use null geodesics, the ``fastest'' causal curves connecting two bulk points.\footnote{We mean the term ``bulk points'' in terms of their coordinates on a constant-time slice.} This is the path of the hare, who tries to win a race by moving as fast as possible on the shortest path. We want to make sure that a signal---the tortoise---makes its trip along the boundary in the shortest time and beats the hare. To remedy (2), we need to quantify the (non-)minimality of encoding intervals in a tractable way.

\subsection{Null Geodesics}

\paragraph*{Explicit parametrization.} The details of the first task are relegated to Appendix \ref{sec:Appendix_A}, where we show that the geodesic equations in $\up{AdS}_3$ are solved in $(t,\g,\th)$ coordinates by
\begin{align}
\label{eqn:result1}
t(\t) = \t, \qquad
\g(\t) = \tan^{-1}\p{\f{\sqrt{1-\ell^2} \tan\t}
{\sqrt{1 + \ell^2 \tan^2\t}}}, \qquad
\th(\t) = \tan^{-1}\p{\ell \tan\t}.
\end{align}
Here $\ell \in [-1,1]$ plays the role of the angular momentum of $p$. Since $\g(\t)$ and $\th(\t)$ are singular at $\t = \f{\pi}{2}$, we extend their definitions ``by hand,'' being careful about the branch cut choices for the inverse tangent. The geodesics are parametrized piecewise:
\begin{align}
\label{eqn:result2}
\g_*(\t) = \begin{cases}
\g(\t), & \t \in [0,\f{\pi}{2}], \\
\g(\pi - \t), & \t \in [\f{\pi}{2}, \pi];
\end{cases} \qquad
\th_*(\t) = \begin{cases}
\th(\t), & \t \in [0,\f{\pi}{2}], \\
\pi - \th(\pi - \t), & \t \in [\f{\pi}{2}, \pi].
\end{cases}
\end{align}

\paragraph*{Conformal structure.}
We can gain some intuition by looking at the Penrose diagram of $\up{AdS}_3$ \cite{Penrose:1962ij}. If we define a shifted coordinate $\g \too \bar{\g} = \g + \f{\pi}{2}$, the AdS metric (\ref{eqn:AdS_metric}) becomes
\begin{align}
\label{eqn:einstein_universe}
\dd s^2_{\up{AdS}_3} = \f{1}{\cos^2\bar{\g}} 
\p{-\dd t^2 + \dd\bar{\g}^2 + \sin^2\bar{\g}\, \dd\th^2} \sim
-\dd t^2 + \dd \W_2^2 = \dd s^2_{M},
\end{align}
where $\sim$ denotes conformal equivalence. This provides a check that AdS is conformal to a patch of the Einstein static universe $M = \R \cross S^2$ \cite{Wald:GR}. It also shows how the radial direction of $\up{AdS}_3$ is folded up into an angular direction of $S^2$ upon conformal rescaling. Finally, it gives a clearer understanding of the null geodesics of $\up{AdS}_3$, which must be the same as those of $M$. Since $M$ has a product metric, the $t$ coordinate of its geodesics is independent of its $(\bar{\g},\th)$ coordinates. The former is given by $t(\t) = E\t + t_1$, while the latter is determined by
\begin{align}
\cot\bar{\g} = a \cos(\th - \th_1) \iff \tan\g = -a \cos(\th - \th_1).
\end{align}

Here $E,\, t_1,\, a$, and $\th_1$ are integration constants; we set $t_1 = \th_1 = 0$ and $E = 1$ to match the analysis in Appendix \ref{sec:Appendix_A}. These geodesics are great circles in $S^2$, so they must meet the equator at diametrically opposite points. Converting back from $\g$ to the standard radial coordinate $r$, we find a new parametrization of the null geodesics:
\begin{align}
r = r(\th) = \sec\g - \tan\g = a \cos\th + \sqrt{1 + a^2 \cos^2\th}.
\end{align}
This curve describes a circular arc intersecting the unit circle at opposite points. Indeed, one may check that $(x - a)^2 + y^2 = 1 + a^2$ for $x = r \cos\th$ and $y = r \sin\th$. Here $a$ plays the role of $\ell$ by controlling how close the geodesic comes to the boundary, thereby giving the bulk point angular momentum. The point of closest approach is at $(r_{\up{min}},\th_{\up{min}}) = (\sqrt{1 + a^2} - a,\, \pi)$, when $a>0$. Thus both spacelike and null geodesics in $\up{AdS}_3$ are circular arcs, the former orthogonal to the boundary and the latter passing through diametrically opposite points.

\subsection{Non-Minimal Encoding}
\label{subsec:non-minimal}

\paragraph*{Gems and mimic points.}
Observe that every geodesic of $D^2$ is uniquely labeled by its ``deepest'' bulk point. We call this point the \it{gem} of the geodesic. Suppose that $p_0 = (\g_0, \th_0)$ is the gem of some geodesic in $D^2$, and let $p = (\g,\th)$ be an arbitrary point on the same geodesic. Then $p_0$ and $p$ may both be encoded by the same boundarry interval, but $p_0$ is encoded minimally while $p$ may not be: it is as if $p$ is pretending to be at $p_0$.

\begin{figure}[H]
\centering
\includegraphics[trim = 120 230 95 215, clip,
width=0.57\textwidth]{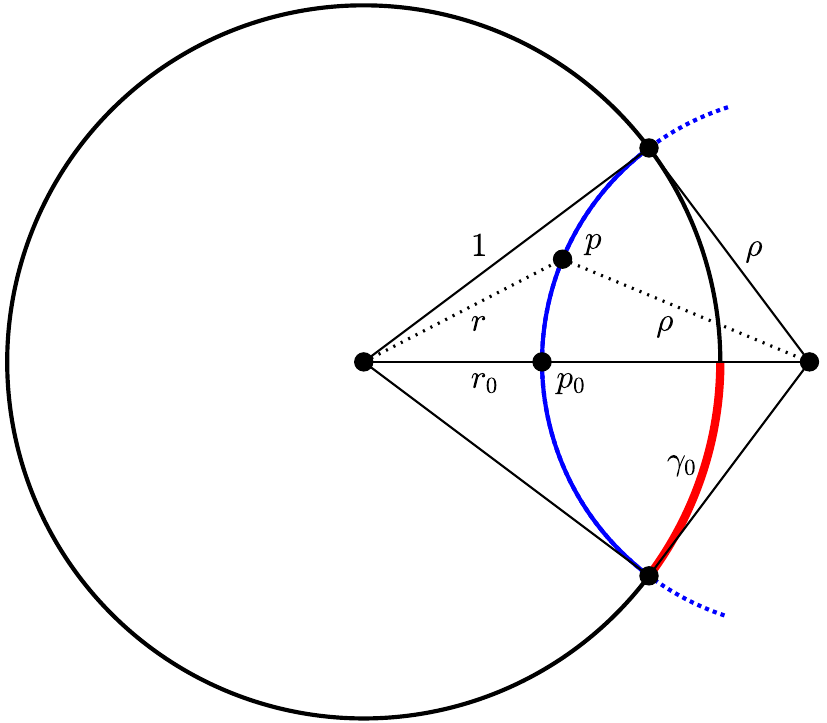}
\caption{Coordinates of a point lying on a geodesic.}
\label{fig:geometry_2}
\end{figure}

As shown in Fig. \ref{fig:geometry_2}, the coordinates of $p$ and $p_0$ are related by the law of cosines:
\begin{align}
\rho^2 = r^2 + (\rho + r_0)^2 - 2r(\rho + r_0) \cos(\th_0 - \th).
\end{align}
We divide through by $(\rho + r_0)^2$ and recognize $\sin\g_0$ from (\ref{eqn:trig}). After some algebra, we obtain
\begin{align}
\sin^2\g_0 &= r^2 \cos^2\g_0 + 1 - 
2r \cos\g_0 \cos(\th_0 - \th) \implies 
\cos\g_0 = {\cos(\th_0-\th)} \cos\g.
\end{align}

We introduce $\a \equiv \th_0 - \th \in [-\f{\pi}{2}, \f{\pi}{2}]$ to measure the degree of encoding asymmetry. We also change notation $p_0 \too p_{\a} = (\g_{\a}, \th_{\a})$ to emphasize $p_{\a}$ as a ``mimic'' point, disheveled from $p$ by angle $\a$. This lets us pretend that the bulk encoding is always minimal, as long as we use $(\g_{\a}, \th_{\a})$ instead of $(\g,\th)$. In this notation, our results are
\begin{align}
\label{eqn:mimic}
\cos\g_{\a} = \cos\a \cos\g, \qquad \th_{\a} = \a + \th, \qquad \a \in \qty[-\tfrac{\pi}{2}, \tfrac{\pi}{2}].
\end{align}

\paragraph*{Locus of gems.}
The discussion above begs a natural question: where can a bulk point pretend to hide? More precisely, what is the locus of gems of geodesics passing through a fixed point $p$ with coordinates $(\g,\th)$? It is just the set of points with coordinates in
\begin{align}
G_p \equiv \qty{(\g_{\a},\th_{\a}) \in \qty[0,\tfrac{\pi}{2}] \cross S^1 \,\Big|\, \cos\g_{\a} = \cos(\th_{\a} - \th) \cos\g}.
\end{align}
The defining equation of $G_p$ is one characterization of its shape; another comes from a conversion of $(\g_{\a}, \th_{\a})$ into polar $(r_{\a},\th_{\a})$ and then into Cartesian $(x,y)$ coordinates. We may use the rotational invariance of $D^2$ to set $\th = 0$ for the moment; then, using (\ref{eqn:trig}), we have
\begin{align}
\f{2r_{\a}}{1 + r_{\a}^2} = \cos\th_{\a} \cos\g &\implies
\k r_{\a}^2 = (1 + r_{\a}^2) r_{\a} \cos\th_{\a} \implies \notag \\
&\implies x^3 + xy^2 - \k (x^2 + y^2) + x = 0, \quad
\k \equiv \f{2}{\cos\g} \in [2,\infty).
\end{align}
So as a curiosity, the set of bulk points encoded by the same boundary interval---the points as which $p$ can masquerade for the purposes of causality---is a cubic curve in $D^2$.

\section{Analysis of Causality}
\label{sec:caus}

\begin{figure}[H]
    \centering
    \begin{minipage}{.5\textwidth}
        \centering
        \includegraphics[trim = 140 230 120 215, clip,
        width=.95\linewidth]{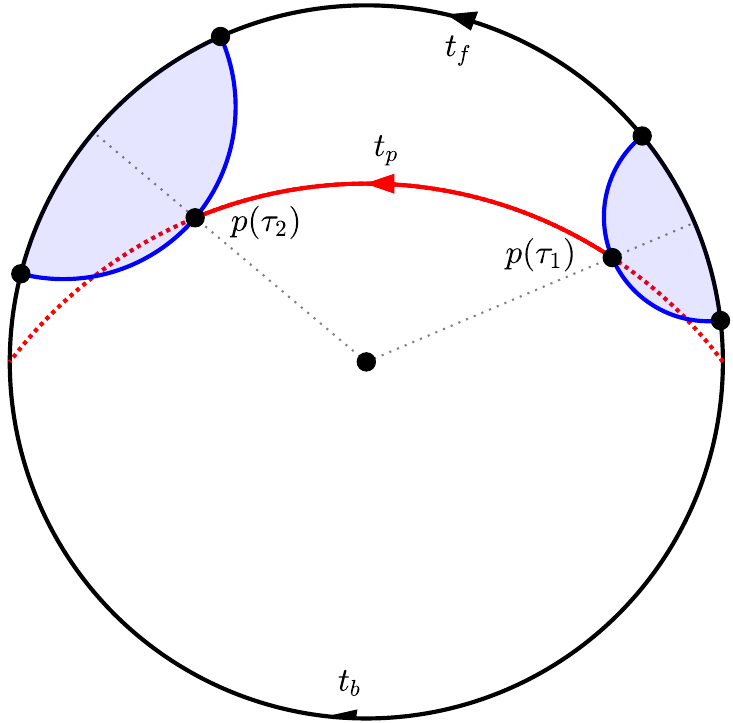}
    \end{minipage}%
    \begin{minipage}{0.5\textwidth}
        \centering
        \includegraphics[trim = 140 230 120 215, clip,
        width=.95\linewidth]{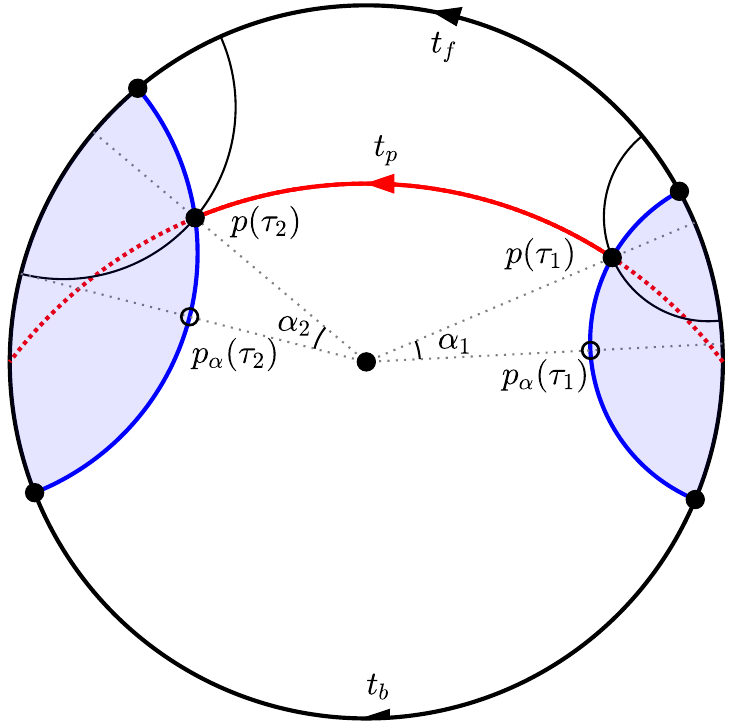}
    \end{minipage}
    \caption{Null geodesic motion: minimal (left, $\a(\t) \equiv 0$) and non-minimal (right, $\a_1,\a_2 \neq 0$) encodings. Minimal encodings (thin curves) and gems of $p$ (open circles) are also shown.}
    \label{fig:geodesic_causality}
\end{figure}

\paragraph*{Physical setup.} We aim to prove that for motion along null geodesics, and hence along any causal curve, the tortoise on the boundary always beats the hare in the bulk. The latter, a freely falling massless observer $p \in \up{AdS}_3$, moves from $p(\t_1)$ to $p(\t_2)$ along a curve parametrized by $p(\t) = (\t, \g_*(\t), \th_*(\t))$ as in (\ref{eqn:result1}--\ref{eqn:result2}). Because $t(\t) = \t$, its trip takes time $t_p = \t_2 - \t_1$. Meanwhile, the travel time of the signals depends on the encoding of $p$ on a boundary region subtended by a spacelike geodesic passing through it. This region may encode $p(\t)$ non-minimally; as shown in Fig. \ref{fig:geodesic_causality}, one way to describe the encoding is to let $\a(\t)$ be the angle by which $p(\t)$ is disheveled from the gem of its encoding geodesic.

\paragraph*{Basic strategy.}
We endeavor to show that at least one boundary signal beats the bulk point as it travels along null geodesics. In other words, we want to prove\footnote{Here and below, (e.g.) $\g_{\a}(\t)$ stands for $\g_{\a(\t)}(\t)$. In particular, for minimal encoding, $\g_0(\t) = \g_*(\t)$.} that
\begin{align}
\label{eqn:time_delay}
\Delta t^{\a}(\t_1, \t_2) &\equiv 
\max\{\Delta t_f^{\a}, \Delta t_b^{\a}\} \equiv
\max\{t_p - t_f^{\a},\, t_p - t_b^{\a}\} = \notag \\ &=
\big( \t_2 - \t_1 \big) - \min\{t_f^{\a},\, t_b^{\a}\} \geq 0,
\end{align}
where $\Delta t_f^{\a} \equiv t_p - t_f^{\a}$ and $\Delta t_b^{\a} \equiv t_p - t_b^{\a}$ are what we might call the forward and backward holographic time delay functions for $\up{AdS}_3$. 

Referring to Fig. \ref{fig:geodesic_causality}, the forward and backward signals' travel times are equal to the angles between the neighboring edges of the shaded regions:
\begin{align}
\label{eqn:signal_time}
\begin{aligned}
t_f^{\a}(\t_1,\t_2) &\equiv
\big( \g_{\a}(\t_2) + \g_{\a}(\t_1) \big) - 
\big( \th_{\a}(\t_2) - \th_{\a}(\t_1) \big), \\
t_b^{\a}(\t_1,\t_2) &\equiv
\big( \g_{\a}(\t_2) + \g_{\a}(\t_1) \big) +
\big( \th_{\a}(\t_2) - \th_{\a}(\t_1) \big) - 2\pi.
\end{aligned}
\end{align}

The case of minimal encoding, $\a(\t) \equiv 0$, yields in subsection \ref{subsec:4.1} to direct attack by a convexity argument much like the one in the analysis of circular motion. For $\a(\t) \neq 0$, proving directly that $\Delta t^{\a} \geq 0$ for $\a(\t) \neq 0$ is still possible, but is neither elegant nor illuminating. Instead, in subsection \ref{subsec:4.2}, we observe that for any path through the bulk, there is a unique critical encoding $\a_*(\t)$ for which both the forward and backward boundary signals arrive simultaneously with the bulk point: $t_f = t_b = t_p \implies \Delta t \equiv 0$. We then study deviations from this critical encoding, as measured by a new angle $\b(\t)$. In subsection \ref{subsec:4.3}, we re-coordinatize $\up{AdS}_3$ using this critical encoding. This converts the claim of causality preservation into a statement of spherical trigonometry, which completes the proof.

\paragraph*{Miscellaneous comments.}
We begin by restricting the ranges of $\ell$ and $\t_1$. First, while $\ell \in [-1,1]$ labels all null geodesics, we consider only $\ell \in [0,1]$ because the paths labeled by $\ell$ and $-\ell$ have identical time delays. The time delay preserves the rotational $\up{U}(1)$ symmetry of $D^2$, as well as a $\Z_2$ reflection symmetry: fixing the boundary point $p(0)$ breaks the $\up{U}(1)$, and choosing the sign of $\ell$ breaks the remaining $\Z_2$. Second, although $0 \leq \t_1 \leq \t \leq \t_2 \leq \pi$ labels all points on a null geodesic, we consider only $\t_1 \in [0,\f{\pi}{2}]$. Every path from $p(\t_1)$ to $p(\t_2)$ with $\f{\pi}{2} \leq \t_1 < \t_2$ starts over halfway into $D^2$ and lies entirely in one quadrant. The $\f{\pi}{2}$-reflected path $p(\pi - \t_2) \too p(\pi - \t_1)$ lies in the quadrant where $\t_1 < \t_2 \leq \f{\pi}{2}$, and has identical time delay. This choice uses up the inversion symmetry, preserved by $\Delta t$, of our geodesics. Next, by way of notation, let us define the function $f(\t)$ and the quantity $L$ by
\begin{align}
f^2(\t) \equiv 1 + \ell^2 \tan^2\t \geq 1, \qquad
L^2 \equiv 1 - \ell^2 \in [0,1].
\end{align}
Note that the function $f(\t)$ appears in \eqref{eqn:result1}, letting us rewrite it as
\begin{align}
\g(\t) = \tan^{-1}\p{\f{L\tan\t}{f(\t)}}, \qquad \th(\t) = \tan^{-1}\p{\ell \tan\t}.
\end{align}
Finally, it may be checked that $\Delta t_f$ and $\Delta t_b$ are both continuous---moreover, at least $C^1$---in both $\t_1$ and $\t_2$. We will make implicit use of this regularity many times.

\subsection{Causality I: Minimal Encoding}
\label{subsec:4.1}

Our first serious result is that causality is preserved for minimally encoded bulk points moving along null geodesics. For convenience, we will omit the superscript $\a$ and the argument $\t_1$ when they vanish, writing (e.g.) $\Delta t^0(0, \t_2)$ as $\Delta t(\t_2)$. We begin by additionally enforcing $\t_1 = 0$, making the bulk point start its journey from the boundary.

\begin{prop}[boundary to bulk]
For all $\ell \in [0,1]$ and all $\t_2 \in [0,\pi]$, we have $\Delta t_f(\t_2) \geq 0$, while $\Delta t_b(\t_2) \leq 0$. Thus the forward signal always preserves causality, while the backward signal lags behind the bulk point.
\end{prop}

\begin{figure}[H]
\centering
\includegraphics[width=0.8\textwidth]{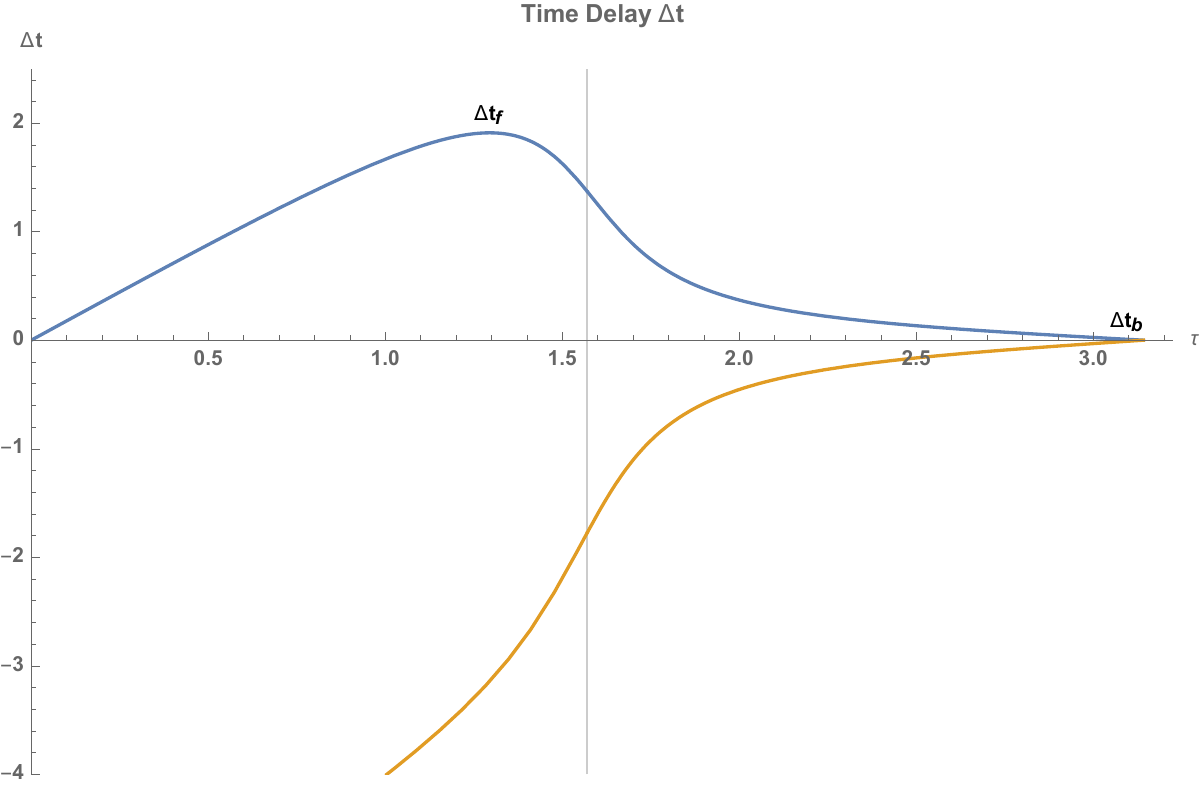}
\caption{Time delay for minimally encoded boundary-to-bulk motion.}
\label{fig:graph_ssbd}
\end{figure}

\begin{mainproof}
We construct the graph of $\Delta t(\t_2)$ in Fig. \ref{fig:graph_ssbd} by establishing the following claims:

\it{Step 1.} $-\Delta t_f(\t_2)$ is convex in $\t_2$ on $[0,\f{\pi}{2}]$. Also, $\Delta t_f(0) = 0$, and $\Delta t_f(\f{\pi}{2}) \geq 0$.

\it{Step 2.} $\Delta t_f(\t_2)$ is decreasing in $\t_2$ on $[\f{\pi}{2},\pi]$. Also, $\Delta t_f(\f{\pi}{2}) \geq 0$, and $\Delta t_f(\pi) = 0$.

\it{Step 3.} $\Delta t_b(\t_2)$ is increasing in $\t_2$ on $[0,\pi]$. Also, $\Delta t_b(0) \leq 0$, and $\Delta t_b(\pi) = 0$.

\noindent Steps 1 and 2 prove that $\Delta t_f(\t_2) \geq 0$, while Step 3 proves that $\Delta t_b(\t_2) \leq 0$.

\bf{Step 1.} To show that $\pdv[2]{}{\t_2} \Delta t_f(\t_2) \equiv \Delta t_f''(\t_2) \leq 0$ for $\t_2 \in [0,\f{\pi}{2}]$ is a direct computation:
\begin{align}
\Delta t_f''(\t_2) = \g''(\t_2) - \th''(\t_2) = 
-\f{\ell \sec^2\t_2}{f^4(\t_2)} 
\Big[2L^2 + \ell L f(\t_2)\Big] \leq 0.
\end{align}
It is also readily checked that $\Delta t_f(0) = 0$ and $\Delta t_f(\f{\pi}{2}) = \tan^{-1}\p{L/\ell} \geq 0$.

\bf{Step 2.} When $\t \in [\f{\pi}{2},\pi]$, we use coordinates $(\g_*, \th_*)$. We check monotonicity for $\Delta t_f(\t_2)$:
\begin{align}
\Delta t_f'(\t_2) = \g_*'(\t_2) - \th_*'(\t_2) = 1 - \f{\ell \sec^2\t_2}{f^2(\t_2)} - \f{L}{f(\t_2)}.
\end{align}
To see that this derivative is never positive, we multiply through by $f^2(\t_2)$. We then use that $\ell \geq \ell^2$ and $f(\t_2) \geq 1$ to bound the last two terms:
\begin{align}
f^2(\t_2) - \ell \sec^2\t_2 - L f(\t_2) \leq 1 + \ell^2\tan^2\t_2 - \ell^2\sec^2\t_2 - L = L^2 - L \leq 0.
\end{align}
It is also direct to check that $\Delta t_f(\pi) = 0$.

\bf{Step 3.} We check monotonicity separately for $\Delta t_b(\t_2)$ on $[0,\f{\pi}{2}]$ and on $[\f{\pi}{2},\pi]$:
\begin{align}
\Delta t_b'(\t_2) = \begin{cases}
1 + \f{\ell \sec^2\t_2}{f^2(\t_2)} + \f{L}{f(\t_2)}, 
& \t_2 \in [0, \f{\pi}{2}], \\
1 + \f{\ell \sec^2\t_2}{f^2(\t_2)} - \f{L}{f(\t_2)},
& \t_2 \in [\f{\pi}{2}, \pi].
\end{cases}
\end{align}
For $\t_2 \in [0,\f{\pi}{2}]$, the nonnegativity of $\Delta t_b'(\t_2)$ is manifest. Meanwhile, for $\t_2 \in [\f{\pi}{2},\pi]$, we have
\begin{align}
\Delta t_b'(\t_2) = -\Delta t_f'(\t_2) + 2 - \f{2L}{f(\t_2)} 
\geq 2 - 2 = 0,
\end{align}
where we have used $\Delta t_f'(\t_2) \leq 0$ (Step 2) and $\f{L}{f(\t_2)} \leq 1$. It is once again straightforward to see that $\Delta t_b(0) = 2\pi > 0$ and $\Delta t_b(\pi) = 0$. This completes the proof.
\end{mainproof}

The content of the proof is visualized in Fig. \ref{fig:graph_ssbd}, where monotonicity and convexity are apparent. Many of the inequalities above are saturated at $\ell = 0$, which describes the radial plunge trajectory of subsection \ref{subsec:radial}. In this case, $\Delta t_f(\t_2)$ is a straight line for $\t_2 \in [0,\f{\pi}{2}]$, and abruptly goes to zero at $\t_2 = \f{\pi}{2}$. Similarly, $\Delta t_b(\t_2)$ is a straight line that reaches zero at $\t_2 = \f{\pi}{2}$ and stays at that value until $\t_2 = \pi$. And when $\ell = 1$, the ``bulk'' point moves on the boundary together with the forward signal, so in that case $\Delta t_f \equiv 0$. 

We now remove the restriction that $\t_1 = 0$; the symmetry argument below shows that starting from the bulk poses no danger to causality.

\begin{prop}[bulk to bulk]
For all $\ell \in [0,1]$, $\t_1 \in [0,\f{\pi}{2}]$, and $\t_2 \in [\t_1,\pi]$, we have $\Delta t_f(\t_1, \t_2) \geq \Delta t_f(0, \t_2) \geq 0$, while $\Delta t_b(\t_1, \t_2) \leq \Delta t_b(0, \t_2) \leq 0$. Thus moving into the bulk only drives $\Delta t_f$ and $\Delta t_b$ apart, and $\Delta t_f$ continues to preserve causality.
\end{prop}

\begin{mainproof}
Observe that we can rewrite the time delay $\Delta t(\t_1, \t_2)$ of (\ref{eqn:time_delay}) in terms of $\Delta t(0,\t_2)$:
\begin{align}
\begin{aligned}
&\Delta t_f(\t_1, \t_2) = \Delta t_f(\t_2) + \d t_f(\t_1), 
& \d t_f(\t_1) = \g_*(\t_1) + \th_*(\t_1) - \t_1, \\
&\Delta t_b(\t_1, \t_2) = \Delta t_b(\t_2) + \d t_b(\t_1),
&\d t_b(\t_1) = \g_*(\t_1) - \th_*(\t_1) - \t_1.
\end{aligned}
\end{align}
Consider the parameter reversal $\t_1 \too \tilde{\t}_1 \equiv \pi - \t_1$, and note that $\t_1 \in [0,\f{\pi}{2}] \implies \tilde{\t}_1 \in [\f{\pi}{2},\pi]$. It follows from the definitions (\ref{eqn:result2}) of $\g_*$ and $\th_*$ that $\d t_f(\t_1) = \Delta t_f(\tilde{\t}_1)$ and $\d t_b(\t_1) = \Delta t_b(\tilde{\t}_1)$:
\begin{align}
\begin{aligned}
\Delta t_f(\tilde{\t}_1) &= \pi - \t_1 + \g_*(\pi - \t_1) - \th_*(\pi - \t_1) \qquad \;
= -\t_1 + \g(\t_1) + \th(\t_1) = \d t_f(\t_1), \\
\Delta t_b(\tilde{\t}_1) &= \pi - \t_1 + \g_*(\pi - \t_1) + \th_*(\pi - \t_1) - 2\pi = -\t_1 + \g(\t_1) - \th(\t_1) = \d t_b(\t_1).
\end{aligned}
\end{align}
So by the previous proposition, the $\delta t$ contributions only strengthen causality preservation:
\begin{align}
\begin{aligned}
\Delta t_f(\t_1,\t_2) &= \Delta t_f(\t_2) + 
\Delta t_f(\tilde{\t}_1) \geq \Delta t_f(\t_2) \geq 0, \\
\Delta t_b(\t_1,\t_2) &= \Delta t_b(\t_2) + 
\Delta t_b(\tilde{\t}_1) \leq \Delta t_b(\t_2) \leq 0.
\end{aligned}
\end{align}
Thus by moving into the bulk ($\t_1 > 0$), we gain additional time delay relative to the case where we start on the boundary ($\t_1 = 0$), and so causality gets easier to preserve.
\end{mainproof}

What happens if the encoding does not stay minimal? Even if we allow $\a(\t)$ to wiggle around while the bulk point moves, the time delay $\Delta t^{\a}$ depends only on the initial and final encoding angles, $\a_1 \equiv \a(\t_1)$ and $\a_2 \equiv \a(\t_2)$. If $\a_2 \leq \a_1$, then the arclength of either the forward or the backward signal must get shorter as the other one becomes longer. The boundary trip is now more efficient, and causality becomes yet easier to preserve: $t_{\up{bdy}}^{\a} = \min\{t_f^{\a}, t_b^{\a}\} \leq \min\{t_f^0, t_b^0\} = t_{\up{bdy}}^0$, so $\Delta t^{\a} \geq \Delta t^0 \geq 0$. Therefore without loss of generality we may take $\a_1 \leq \a_2$. But even under this assumption, a direct analysis of $\Delta t^{\a}(\t_1, \t_2)$ quickly becomes very technical.

\subsection{Causality II: Critical Encoding}
\label{subsec:4.2}

\begin{figure}[H]
\centering
\includegraphics[trim = 125 230 104 215, clip,
width=0.57\textwidth]{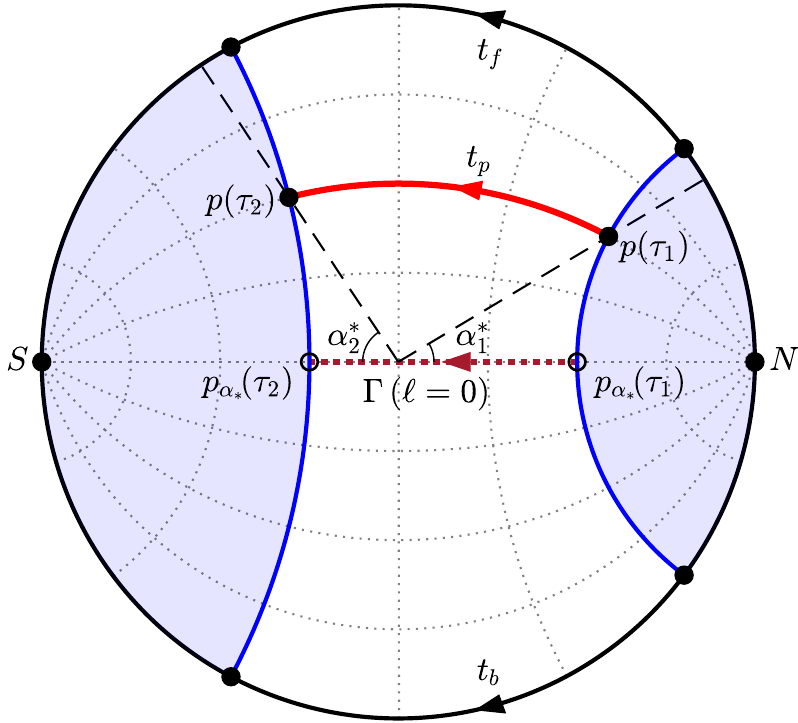}
\caption{The critical encoding and the geographic $(\bar{\psi},\phi)$ coordinate system it begets, shown by the dotted curves and described in subsection \ref{subsec:4.3}. The dashed segment, along $\G$, tracks the gems $p_{\a}(\t)$. We denote $\a_*(\t_1) \equiv \a^*_1$ and $\a_*(\t_2) \equiv \a^*_2$ for shorthand.}
\label{fig:critical_encoding}
\end{figure}

Consider the encoding symmetric about the null geodesic $\G$ with $\ell = 0$, i.e. the line between $N = p(0)$ and $S = p(\pi)$. As shown in Fig. \ref{fig:critical_encoding}, we encode every bulk point $p \in D^2$ on the boundary interval\footnote{Recall that we are working in a fixed time-slice. We will abuse language, saying (e.g.) that bulk points in $D^2$ are encoded on boundary intervals, that (null and spacelike) geodesics are curves in $D^2$, and so on.} subtended by the geodesic through $p$ whose gem lies on $\G$. The symmetry of the encoding about $\G$ ensures that both the forward and backward boundary signals cover the same arclength, so they arrive simultaneously: $t_f = t_b$.

We claim that this encoding, whose angle to the minimal encoding we call $\a_*(\t)$, is always critical: $\Delta t^{\a_*}(\t_1, \t_2) \equiv 0$. To see this, recall from subsection \ref{subsec:non-minimal} that the time delay $\Delta t^{\a}$ of any path $p(\t)$ in the bulk, encoded at angle $\a(\t)$, is equal to the time delay $\Delta t^0$ of the \it{minimally} encoded path of the \it{mimic} point $p_{\a}$, which follows the gems of the geodesics encoding $p$. In our case, the gems $p_{\a}$ lie on $\G$ and describe radial infall. We studied their the minimal encoding in subsection \ref{subsec:radial} and found that $\Delta t^0$ vanishes identically for $p_{\a}$; we will conclude that $\Delta t^{\a_*} \equiv 0$ for $p$ as well. It is deviations from $\a_*$ to which we turn next. 

\begin{figure}[H]
    \centering
    \begin{minipage}{.5\textwidth}
        \centering
        \includegraphics[trim = 140 230 120 215, clip,
        width=.95\linewidth]{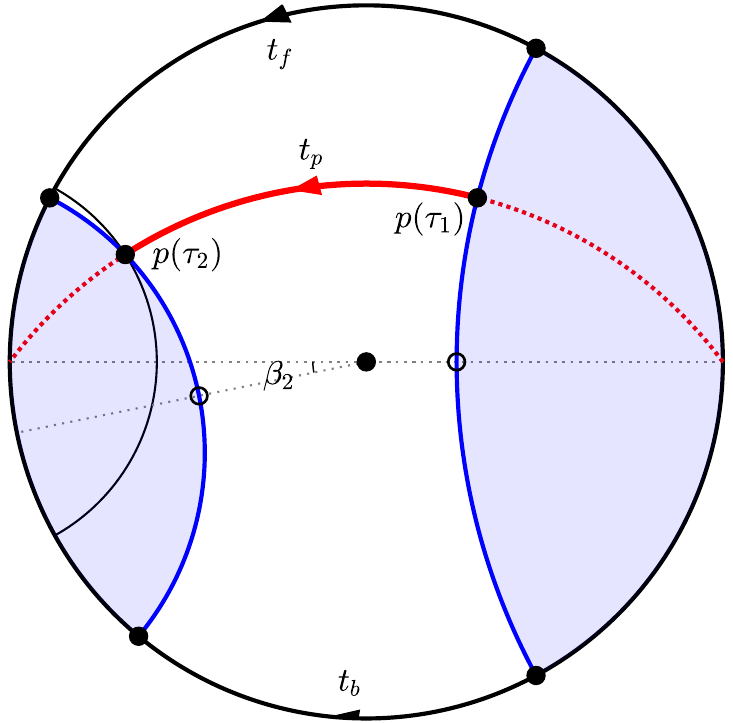}
    \end{minipage}%
    \begin{minipage}{0.5\textwidth}
        \centering
        \includegraphics[trim = 140 230 120 215, clip,
        width=.95\linewidth]{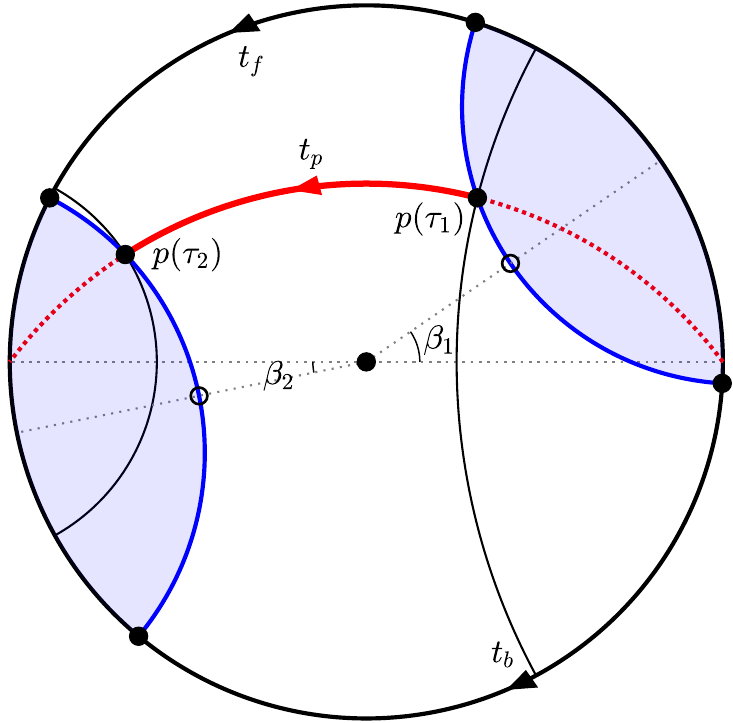}
    \end{minipage}
    \caption{Non-critical encodings. Left: $\b_1 = 0$, $\b_2 \neq 0$. Right: $\b_1 \neq 0 \neq \b_2$.}
    \label{fig:non_critical}
\end{figure}

To that end, we introduce the angle $\b(\t)$ by which the encoding fails to be critical.\footnote{More precisely, $\b(\t)$ is the signed angle pointing from $\G$ to the gem of the geodesic that encodes $p(\t)$. We say that $\b > 0$ if it points counterclockwise. This convention also applies to the angles $\a_*(\t)$ in Fig. \ref{fig:critical_encoding} and helps to clarify the result $\a_* \sim -\th_*$ below.} As with $\a(\t)$, the time delay, now labeled $\Delta t^{\b}(\t_1, \t_2)$, depends only on $\b_1 \equiv \b(\t_1)$ and $\b_2 \equiv \b(\t_2)$, as shown in Fig. \ref{fig:non_critical}. If $\b_1 = 0$, then $\b_2 \neq 0$ will lengthen one of the boundary travel times and shorten the other, relative to the critical travel time $t_f = t_b = t_p$. The same is true if $\b_2 = 0$ and $\b_1 \neq 0$. Thus if either endpoint of the bulk point's path is critically encoded, causality is preserved:
\begin{align}
\begin{aligned}
\Delta t(\b_1 = 0 \neq \b_2) \geq \Delta t(\b_1 = 0 = \b_2) = 0, \\
\Delta t(\b_1 \neq 0 = \b_2) \geq \Delta t(\b_1 = 0 = \b_2) = 0.
\end{aligned}
\end{align}

It remains to consider the general case, $\b_1 \neq 0 \neq \b_2$. Before we do so, a few comments are in order. First, we can assume without loss of generality that $\b_1$ and $\b_2$ have the same sign; otherwise, they work together to increase $\Delta t$ even further above its value when either $\b_1$ or $\b_2$ vanishes. Moreover we can take both to be positive, since the symmetry of the critical encoding guarantees that the time delay for an encoding with $\b_1, \b_2 < 0$ is equal to the time delay for the encoding with $-\b_1$ and $-\b_2$. Second, because a generic encoding geodesic intersects the critical one away from $\G$, we no longer have $\b_1 \leq \b_2$ in the same way we had $\a_1 \leq \a_2$ above. Nevertheless, by definition $\a(\t) = \a_*(\t) + \b(\t) \in [-\f{\pi}{2},\f{\pi}{2}]$. This condition affords us the weaker constraint $\a_1 + \b_1 \leq \a_2 + \b_2$. 

To summarize, $\a$ measures the non-minimality of an encoding, $\b$ measures its non-criticality, and $\a_*$ measures how non-minimal the critical encoding is. This clarifies why radial infall along $\G$ is special: it is both minimal and critical, $\a(\t) = \a_*(\t) = \b(\t) \equiv 0$. As we proceed to the general case, it is the $\b$-perspective that will become ``critical'' to us.

\subsection{Causality III: The General Case}
\label{subsec:4.3}

\paragraph*{New coordinates.}
We still lack a convenient coordinate system in which to describe deviations $\b(\t)$ from criticality. Such a coordinate system will consist of the null and (critical) spacelike geodesics, as shown in Fig. \ref{fig:critical_encoding}. Every bulk point is labeled uniquely by its null geodesic (how far ``up'' it is) and by its critical encoding (how far ``in'' it is). This coordinate ruling is orthogonal, and viewed from the side, it resembles the meridians and parallels on a globe. Indeed, we now pass from the faraway imagination of $\up{AdS}_3$ to the close reality of our Earth. The idea is to take advantage of the fact that the angular momentum $\ell$ is constant along each null geodesic, while the length $\g_{\a_*}$ is constant along each critical spacelike geodesic. For $\t \in [0,\f{\pi}{2}]$, null geodesics are described by (\ref{eqn:result1}). We may rewrite the equation for $\th(\t)$ as $\tan\t = \f{1}{\ell} \tan\th(\t)$ and substitute it into the equation for $\g(\t)$; this yields
\begin{align}
\label{eqn:meridians}
\tan \g(\t) = \f{L}{\ell} \sin \th(\t) \implies
\f{\tan\g}{\sin\th} = \f{L}{\ell} \equiv \tan\phi = \up{constant}.
\end{align}
Meanwhile, the critical spacelike geodesics are characterized by
\begin{align}
\label{eqn:parallels}
\cos \g_{\a_*}(\t) = \cos \a_*(\t) \cos \g(\t) = 
\cos\th \cos\g \equiv \cos\psi = \up{constant},
\end{align}
where we have used (\ref{eqn:mimic}) and the fact that $\a_* = -\th$.  We have introduced the constants $\phi = \tan^{-1}\p{L/\ell} = \g(\f{\pi}{2})$ and $\psi = \g_{\a_*}$: the former measures how close a null geodesic comes to the boundary, and the latter measures the depth of a critical spacelike geodesic's gem. One can check using (\ref{eqn:result2}) that the results above hold for $\t \in [\f{\pi}{2}, \pi]$ as well as for $\t \in [0, \f{\pi}{2}]$. 

To complete the coordinate change, we solve (\ref{eqn:meridians}--\ref{eqn:parallels}) for $\g$ and $\th$ in terms of $\psi$ and $\phi$ and rewrite the $\up{AdS}_3$ metric (\ref{eqn:AdS_metric}) in the new coordinates $(t,\psi,\phi)$:
\begin{align}
\label{eqn:sphere}
\begin{aligned}
\dd s^2_{D^2} &=
\f{1}{\sin^2 \phi\, \sin^2\psi} 
\p{-\dd t^2 + \dd \psi^2 + \sin^2\psi \, \dd\phi^2} = \\ &=
\f{1}{\sin^2 \phi\, \cos^2\bar{\psi}} 
\p{-\dd t^2 + \dd \bar{\psi}^2 + 
\cos^2\bar{\psi} \, \dd\phi^2}.
\end{aligned}
\end{align}
In terms of $(\psi, \phi)$, the spatial part of the metric is conformal to a patch of the round sphere with an adapted north and south pole on the boundary of $D^2$. Meanwhile, shifting to $\bar{\psi} \equiv \psi - \f{\pi}{2}$, we find that $\dd s^2$ takes the same form in coordinates $(t,\g,\th)$ as in $(t,\bar{\psi}, \phi)$. In fact, $\bar{\psi}$ is another tortoise coordinate, perhaps more natural for holography than $\g$. In these coordinates, null ($\dd \phi = 0$) and critical spacelike ($\dd \bar{\psi} = 0$) geodesics play the roles of the radial and circular paths from section \ref{sec:illus}, respectively.\footnote{To prove that causality is preserved, it suffices to extend the analysis of radial infall to arbitrary encodings. Equivalently, from Fig. \ref{fig:non_critical}, we must show that the encoding intervals remove more arclength on one side of the boundary circle than they create on the other. But we will not pursue these ideas directly.}

Since our analysis is insensitive to conformal rescaling, the metric (\ref{eqn:sphere}) gives us license to work entirely on $S^2$ and affords a fresh interpretation of the time delay $\Delta t$. Null geodesics become meridians or lines on $S^2$, encoding geodesics become parallels or circles on $S^2$, and $t_p$ is the length of a meridian arc between the parallels encoding the initial and final bulk points. An encoding is critical ($t_p = t_f^{\b = 0} = t_b^{\b = 0}$) if all of the encoding parallels share a center---the north pole---that also lies on the meridian. Meanwhile, non-critical encodings are parallels with an arbitrary center whose location is related to the angle $\b$ in Fig. \ref{fig:non_critical}.

\begin{figure}[H]
    \centering
    \includegraphics[trim = 125 230 100 210, clip,
    width=.57\linewidth]{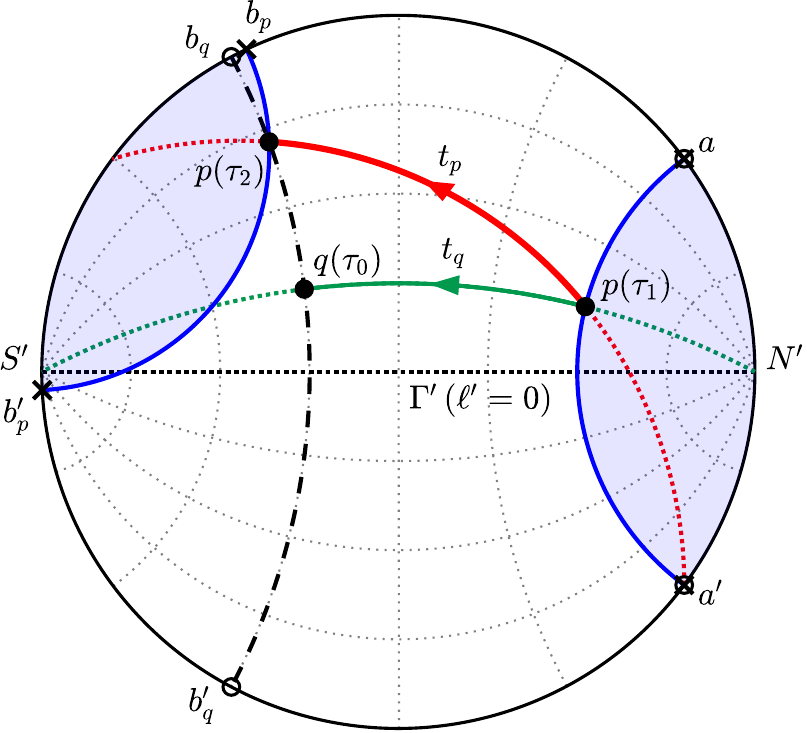}
    \caption{$S^2$ and $D^2$, rotated by angle $\b_1$. The initial encoding, a parallel with center $N'$, is critical for a new null geodesic $q(\t)$, a meridian through $N'$, with constant $\phi'$. The initial ($a$) and final ($b$) signal locations for paths $p$ ($\circ$) and $q$ ($\cross$) are indicated on the boundary.}
    \label{fig:rotated}
\end{figure}

\paragraph*{Proof of the general case.}
To attack the case $\b_1 \neq 0 \neq \b_2$, we rotate the globe (and $\up{AdS}_3$) by angle $-\b_1$ to ``undo'' the non-criticality of the initial encoding. As shown in Fig. \ref{fig:rotated}, we introduce new coordinates $(\psi', \phi')$ adapted to the initial encoding, which becomes the latitude circle $aa'$ of $p(\t_1)$, and whose center becomes the new north pole $N'$. This encoding becomes critical for the family of null geodesics taking the form (\ref{eqn:result1}--\ref{eqn:result2}) in coordinates $(\bar{\psi}',\phi')$, i.e. those with constant $\ell'$ (and hence constant $\phi'$). Among these is the unique null geodesic $q$ passing through $p(\t_1)$. Let us consider a critically encoded bulk point that moves along $q$: it starts at $p(\t_1)$, where it shares the initial encoding $aa'$ with the bulk point $p$, and it arrives at the point $q(\t_0)$ where it orthogonally intersects---and is encoded by---the latitude circle $b_q b_q'$ of $p(\t_2)$. Meanwhile, $p(\t_2)$ itself is encoded by the generic circle $b_p b_p'$, which intersects the parallel $b_q b_q'$ exactly once\footnote{Circles on the sphere actually intersect in two points, but $\up{AdS}_3$ is conformally equivalent to a \it{hemisphere} of $S^2$ where such intersections are unique. A clever way to see this is to perform an inversion about the center of $D^2$ that preserves the boundary circle. The inversion maps circles to circles, takes the inside of $D^2$ to the outside, and preserves the geodesic circles in $D^2$ because it preserves angles. The intersection point inside $D^2$ is then moved to a point outside $D^2$.} at $p(\t_2)$.

\newpage
We show presently that the bulk travel time $t_q$ of $q$ exceeds the boundary travel time $\min\{t_f, t_b\}$ of $p$, but is dominated by the bulk travel time $t_p$ of $p$; this demonstrates that $\min\{t_f, t_b\} \leq t_q \leq t_p$ and completes the proof of causality preservation in the general case. Since $q$ is critically encoded, $t_q$ is equal to both of its attendant signals' travel times, which are the standard polar angles $t_q = \th(a, b_q) = \th(a', b_q')$. But the arcs $b_q b_q'$ and $b_p b_p'$ intersect once and then bow out, so by previous arguments (see Fig. \ref{fig:rotated}) one of the angles $t_f = \th(a, b_p)$ or $t_b = \th(a', b_p')$ will be smaller than $t_q$, while the other will be larger. Thus we have the first part of our claim, $\min\{t_f, t_b\} \leq t_q$. 

\begin{figure}[H]
    \centering
    \includegraphics[width=.57\linewidth]{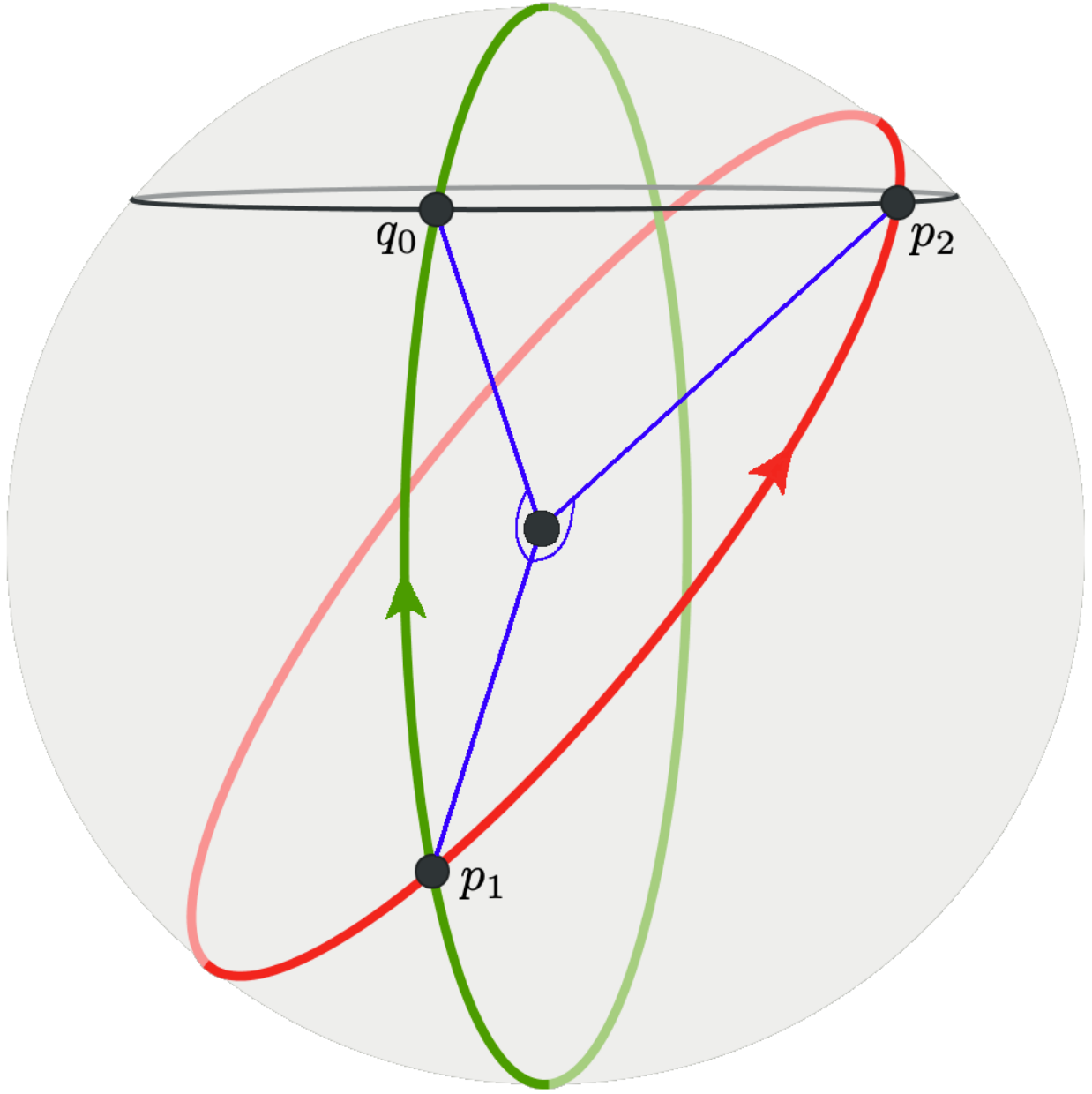}
    \caption{The claim $t_q \leq t_p$, illustrated. We denote $p(\t_1) = p_1$, $p(\t_2) = p_2$, and $q(\t_0) = q$.}
    \label{fig:sphere}
\end{figure}

Meanwhile, the claim that $t_q \leq t_p$ can be translated into the situation shown in Fig. \ref{fig:sphere}: two meridians $p$ and $q$ on (a hemisphere of) $S^2$ intersect at $p_1 = p(\t_1)$, and a circle perpendicular to $q$ intersects both meridians at $p_2 = p(\t_2)$ and $q_0 = q(\t_0)$, respectively. We must prove that the arc $t_q = p_1 q_0$ is shorter than the arc $t_p = p_1 p_2$. Equivalently, viewing $S^2 \subset \R^3$ as a collection of unit vectors, the angle $\th(p_1, q_0)$ made by the vectors $p_1$ and $q_0$ must be smaller than the angle $\th(p_1, p_2)$. 

We begin by fixing the spherical coordinates $(r,\th,\phi)$ of our points and their ranges:
\begin{align}
p_1 = (1, \th_p, 0), \quad p_2 = (1, \th_q, \phi), \quad
q_0 = (1, \th_q, 0); \qquad \th_p, \th_q \in [0,\pi], \quad \phi \in [-\tfrac{\pi}{2}, \tfrac{\pi}{2}].
\end{align}
The angles $\th(p_1, q_0)$ and $\th(p_1, p_2)$ are computed from their dot products, which are found using the Jacobian of the transformation between Cartesian and spherical coordinates:
\begin{align}
(r_1, \th_1, \phi_1) \cdot (r_2, \th_2, \phi_2) = r_1 r_2 \big(
\sin\th_1 \sin\th_2 \cos(\phi_2 - \phi_1) + \cos\th_1 \cos\th_2 \big).
\end{align}
We compute the difference in the arclengths $t_q = \th(p_1, q_0)$ and $t_p = \th(p_1, p_2)$ explicitly:
\begin{align}
\th_{pq}(\phi) &\equiv t_p - t_q =  \th(p_1, p_2) - \th(p_1, q_0) = 
\cos^{-1}\p{p_1 \cdot q_0} - \cos^{-1}\p{p_1 \cdot p_2} = \notag \\ &=
\cos^{-1}\Big(\sin\th_p \sin\th_q \cos\phi + \cos\th_p \cos\th_q\Big) -
\p{\th_p - \th_q}.
\end{align}
It is straightforward to show that $\th_{pq}(\phi)$ is increasing for $\phi \geq 0$:
\begin{align}
\pdv{\th_{pq}}{\phi} = \f{\sin\th_p \sin\th_q \sin\phi}{\sqrt{1 - \p{p_1 \cdot p_2}^2}} \geq 0.
\end{align}
To wit, $1 - (p_1 \cdot p_2)^2 \geq 0$ since $p_1$ and $p_2$ are unit vectors. Also, $\sin\th_p$ and $\sin\th_q$ are nonnegative whenever $\th_p$ and $\th_q$ lie between $0$ and $\pi$. And finally, $\sin\phi \geq$ for $\phi \in [0, \f{\pi}{2}]$.

But now we are done: observe that $\th_{pq}$ is an even function of $\phi$ (due to the symmetry of the latitude circle), vanishes at $\phi = 0$ (where the meridians $p$ and $q$ become identical), and increases for $\phi \geq 0$. By symmetry, $\th_{pq}$ is always nonnegative and attains its minimum value at $\phi = 0$. This shows that the orthogonal meridian $q$ provides the shortest path from $p_1$ to the parallel at latitude $\th_q$, and that moving further out on the latitude circle (changing $\phi$, or equivalently varying $\b_2' = \b_2 - \b_1$) can only give longer bulk travel times.

\paragraph*{Summary and a generalization.}
Let us recapitulate. By adapting the $(\bar{\psi},\phi)$ coordinate system to either the initial or final encoding of the bulk point, we discovered coordinates where, fixing one encoding interval, there is a unique encoding of the other point for which all boundary and bulk times of flight are the same. (We fixed the initial encoding purely for convenience.) These coordinates have a north pole at the midpoint of the initial encoding interval. When we toggle the angle of the final interval about the point of arrival, one boundary time becomes shorter and the other gets longer. Unless the bulk point starts and arrives on the same meridian, its time of flight exceeds the travel time on the boundary. 

This argument also proves causality in $\up{AdS}_d$, so long as bulk points are encoded in spherical regions of the boundary. Latitude circles are replaced by sphere domes in the Poincar\'e ball $D^{d-1}$, generated by the geodesics of $D^{d-1}$ tangent to any point of the surface in the interior. The time delays for starting and landing on these domes is uniform in these coordinates: the meridians are infalling coordinates relative to the north pole, just as in $\up{AdS}_3$. The domes for different sphere encodings of the arrival point are determined by the plane of tangency at the arrival point. Toggling the plane of arrival against the critical plane yields two transverse hyperplanes that share a common geodesic as long as the dimension $d-1$ of the ball is at least 3. (Such geodesics degenerate to a point in $D^2$.) More spherical angles may be toggled, but the result is the same: part of the dome at the boundary necessarily moves closer to the original encoding along at least one geodesic.

\section{Discussion and Conclusions}
\label{sec:disc}

In this paper, we introduced an apparent conflict between the bulk and boundary perspectives on causality: a bulk point traveling through empty $\up{AdS}_3$ seems to drag its entanglement wedge along with it faster than light can move on the boundary. We demonstrated in section \ref{sec:illus} how this paradox is resolved when the bulk point travels along circular and radial trajectories. Along the way, we introduced a convenient coordinate $\g$ that measures the size of the point's encoding interval, its location in the bulk, and the nonlocality of its encoding on the boundary. It also turns out to be a tortoise coordinate on $\up{AdS}_3$, and its construction is perhaps our main technical innovation. We went on in section \ref{sec:coordinate} to parametrize null geodesics in coordinates ($\g,\th$) and used these coordinates to discuss the minimality of the bulk encoding. These steps prepared us to prove in section \ref{sec:caus} that for null geodesics and hence for all causal bulk paths, the bulk point's travel time always matches or exceeds the travel time of a signal moving along the boundary at the speed of light. The central idea was that every null geodesic has a critical encoding where the bulk point arrives simultaneously with both the forward and backward boundary signals. By re-coordinatizing $\up{AdS}_3$ using these encodings, we converted the study of the time delays of non-critical encodings into simple statements in spherical geometry and completed the proof that causality is preserved. 

For any two points on a spatial slice of AdS orthogonal to a timelike Killing field, a unique fastest null geodesic connects them. When the entanglement wedges of the two points are adapted to this null geodesic, causality is critical: the time delay between the bulk point and the boundary entanglement wedges vanishes. This means that in AdS, bulk causality is on the verge of violating boundary causality. If causality on the boundary is not violated, then perturbations of the AdS geometry must lead to additional time delay in the bulk. This idea suggests that an extension of the Gao-Wald theorem into the interior of $\up{AdS}$ may follow from holography. Namely, causality in the bulk should be encoded in the bounds imposed on boundary causality by the entanglement wedge.

Given this intuition, one can argue that to ensure this time delay, the metric perturbations must satisfy an energy condition. Such an inequality should take the form of an integral over the null geodesic to leading nontrivial order in perturbation theory, similar to the formulation of the averaged null energy condition (ANEC). That causality is critical suggests that this is a second-order effect: a linearized perturbation of the metric due to gravity can in principle have either sign, and so should the linearized effect. The vanishing of this linear effect would then constrain the equations of motion for the metric in the bulk. It should be noted that a connection between the ANEC and causality has been proven in quantum field theory \cite{Hartman:2016lgu}. Here we do not work with complete null geodesics, but we do have entanglement wedges and Ryu-Takayanagi surfaces playing a role. Our result suggests that there might be a refined form of the ANEC for finite pieces of null geodesics which includes information from the entanglement wedges that connect their endpoints to a boundary. In AdS, these entanglement wedge surfaces are orthogonal to the null geodesic at the points of departure and arrival. The ingredients involved suggest that this is an application of the quantum null energy condition to the AdS perturbations \cite{Bousso:2015wca}, which connects it more directly to the ANEC. It would be very interesting to understand this possibility better. 

Our results are strongly dependent on the details of the spacelike geodesics in \eqref{eqn:AdS_metric0}, which determine the entanglement wedges. It is interesting to generalize to black hole spacetimes, where already there are timelike geodesics describing particles whose angular frequencies exceed unity (see, for example, \cite{Berenstein:2020vlp}). Therefore at least for circular motion, black holes seem to make the causality problem more severe. The resolution must be that the entanglement wedges need to be ``larger,'' in some sense, than in pure $\up{AdS}$. There are also additional complications, like the presence of causal shadows and the fact that the geometry of the entanglement wedges may not be easily calculable. Our analysis of the $\up{AdS}$ geometry, which assumes that null geodesics start and end on the boundary, will generically fail: a black hole can cause some or all of these geodesics to fall into the singularity. Because our explicit tools cannot address these issues, we consider such analysis beyond the scope of the present paper and leave it to future work.

\acknowledgments
D. B. would like to thank J. Simon for discussions. The work of D.B. is supported in part by the Department of Energy under grant DE-SC 0011702.

\appendix

\newpage
\section{Null Geodesics in AdS}
\label{sec:Appendix_A}

To look for null geodesics, let us start by imposing the null-worldline condition:
\begin{align}
\dd s_{\up{AdS}_3}^2 = 
g_{\mu\nu} \dot{x}^{\mu} \dot{x}^{\nu} = 0 \implies
-\dot{t}^2 + \dot{\g}^2 + \cos^2\g\, \dot{\th}^2 = 0.
\end{align}
Here dots are derivatives with respect to an affine parameter $\t$. We seek Killing vector fields (KVFs) that give quantities conserved along the geodesics, but since ``light is conformal,'' conformal KVFs (cKVFs) also yield conserved quantities. The metric is independent of $t$ and $\th$, so $\xi = \partial_{t} = (1,0,0)$ and $\eta = \partial_{\th} = (0,0,1)$ are cKVFs for $\up{AdS}_3$. These give rise to a conserved energy $E \in \R_+$ and angular momentum $\ell \in [-1,1]$:
\begin{align}
-g_{\mu\nu} \xi^{\mu} \dot{x}^{\nu} = \dot{t} \equiv E, \qquad
g_{\mu\nu} \eta^{\mu} \dot{x}^{\nu} = \cos^2\g\, \dot{\th} 
\equiv \ell \implies \dot{\th} = \f{\ell}{\cos^2\g}.
\end{align}
This immediately solves the geodesic equation for $t$ as $t(\t) = E\t + t_0$ and justifies the name $\t$ for the affine parameter. Substituting these conserved quantities into the null condition above, we convert it into an effective Newtonian problem:
\begin{align}
-E^2 + \dot{\g}^2 + \ell \dot{\th} = 0 \implies
\f{1}{2} \dot{\g}^2 + \f{\ell^2}{2 \cos^2\g} = \f{1}{2} E^2.
\end{align}
This ODE is separable, and can be integrated exactly and inverted to find $\g(\t)$:
\begin{align}
\dv{\g}{\t} = \sqrt{E^2 - \f{\ell^2}{\cos^2\g}} \implies \g(\t) = 
\tan^{-1} \p{\pm\f{\sqrt{E^2 - \ell^2} \tan(E(c \pm \t))}
{\sqrt{E^2 + \ell^2 \tan^2(E(c \pm \t))}}}.
\end{align}
We now return to the conservation law for $\ell$ and view it as an ODE for $\th(\t)$. Substituting the solution $\g(\t)$, we integrate the differential equation to find
\begin{align}
\th(\t) = \ell \int \f{\dd\t}{\cos^2\p{\g(\t)}} + \th_0 = 
\tan^{-1} \p{\f{\ell}{E} \tan\p{E(c \pm \t)}} + \th_0.
\end{align}
Here $t_0, c \in \R$ and $\th_0 \in S^1$ are integration constants, while $E \in \R_+$ and $\ell \in [-1,1]$ are conserved quantities. Without loss of generality, we may take $t_0 = \th_0 = 0$ to fix the zero-points of $t(\t)$ and $\th(\t)$. We may also set $E = 1$, since $\g(\t)$ and $\th(\t)$ depend only on the radio $\ell/E$. Finally, the constant $c$ parametrizes how far from the boundary $p$ is at $\t=0$. We set $c=0$ to fix $p \in \partial D^2$ at $\t=0$; if $p$ starts from within the bulk, we will ``start its clock'' at $\t = \t_1 > 0$. These simplifications immediately yield (\ref{eqn:result1}) above and leave $\ell$ as the sole parameter labeling the geodesics. As noted above, we can extend \eqref{eqn:result1} to $\t \geq \f{\pi}{2}$ by reversing the direction of $\t$ and reflecting $\th$ across $\th = \f{\pi}{2}$; this gives \eqref{eqn:result2} directly.

\end{document}